\documentstyle[confproc,named]{article}
\begin{document}
\input{epsf}
\title{Lattice Gas Automata: a discrete model for simulation
       of continuous transport phenomena in
       packages of agricultural products}

\author{R.G.M. v.d. Sman and M.F.M. Janssens \\
	Systems Research \& Development Department, \\
	Agrotechnological Research Institute \\
	p.o. box 17, 6700 AA Wageningen, the Netherlands, \\
	e-mail: R.G.M.van.der.Sman@ato.agro.nl}
\maketitle

\begin{abstract}

A Lattice Gas Automata (LGA) model has been used for simulations of
physical transport phenomena which occur during the conditioning
of packed agricultural products. These phenomena are air flow,
heat and vapour transport. The objective is to use the LGA model for
optimizing packaging design leading to improved quality of the packed
agricultural product.

LGA is a recently developed methodology for simulation
of fluid flow and other transport phenomena. Simulations are done with
lattice gas particles populating a regular lattice. The particles move from
site to site with velocities equal to a lattice spacing per time step.
At lattice sites the particles  can collide with each other.

The results of the LGA model, simulating the heat and vapour transport
processes, are compared with
results from cooling down experiments with packed cut flowers.
The simulation results give
a good approximation of the average behaviour of the packed cut flowers.
LGA seems to be a promising technique for reaching the objective
of the research programme.
\end{abstract}

\section{Introduction}

Packaging is an important factor of agricultural product quality control.
Packages should allow active conditioning of the product and also protect
the product from outside climate changes during storage. Low temperatures
and high air humidity are favourable conditions for the quality of
      fresh agricultural
products. The common method for active conditioning is forced ventilation of
cold air through or alongside the package. Appropriate holes enable
the ventilation of air through the package and still let the package act as
a buffer, protecting the products from outside climatic changes.

The objective of this research programme is to build a simulation model,
which can act as an instrument in the process of optimization
of the packaging design for improved product quality. The input of the
model should be any specific packaging design and agricultural product. The
outcome of the simulation should be the physical behaviour of
the packaged product during active conditioning and storage.
The physical processes relevant to active conditioning
and storage are:
\begin{itemize}
\item   Convective transport of heat and vapour by air flow,
\item   Diffusive transport of heat and vapour in air, and
\item   Heat and vapour transfer from product to air
\end{itemize}
This paper describes
a preliminary model for these transport phenomena. The simulation model
is based on the technique of Lattice Gas Automata.

Lattice Gas Automata (LGA) is  a recently developed methodology for
simulation of fluid flow and other physical transport phenomena.
\cite{Frisch87,Benzi92}. LGA is a special class
of cellular automata. Like cellular automata the problem to be simulated
is mapped on a regular lattice.

The lattice is a fictitious simplified discrete microworld. Space, time,
velocity and the number of particles are all discrete. The positions
of the lattice gas particles are restricted to the lattice sites. At
discrete time steps the particles move to the neighbouring site, accoording
to their velocity. Because of the discreteness of the lattice and time,
the lattice gas particles can only assume a discrete set of velocities
$\{{\bf c}_i\}$, equal to the lattice spacing.
The state of a lattice site is determined by the number
of particles $f_i$ moving with velocity (${\bf c}_i$). At a lattice site
particles may collide with each other. The collisions obey the conservation
laws of mass and momentum.
\begin{figure}[hbt]
\epsfxsize = \hsize
\epsfbox[24 445 582 748]{fig1hex.eps}
\caption{Propagation and collisions of lattice gas
	 particles on a hexagonal lattice}
\end{figure}

In spite of the fact that the dynamics of LGA is similar
to that of the microscopic world of molecules and atoms,
it is expected that LGA is capable to simulate
fluid dynamics at scales of the macroscopic world,
such as turbulent flow. As the Navier Stokes partial differential
equation, by which fluid dynamic phenomena are traditionally
described, LGA is merely a statement of the conservation
laws of mass and momentum. These conservation laws are the only
necessary conditions to get macroscopic fluid dynamic behaviour.
It has been mathematically proven that LGA and the Navier Stokes equation
indeed lead to similar results \cite{Frisch87,Benzi92}.

An example of the dynamics of LGA is shown in figure 1.
This shows colliding particles on
a hexagonal lattice, the simplest two-dimensional lattice showing
fluid dynamic behaviour. At time $t$ the particles collide.
The situations at times $t-1$ and $t+1$ are shown.
The results of the collisions are clearly seen.

Simulations of fluid dynamics with LGA are done with a simple scheme
which is an iteration of two steps: collision and propagation.
The fluid dynamic behaviour is described by the macroscopic parameters,
mass density ($\rho$) and momentum density ($\rho {\bf u}$),
which are derived from the particle distribution function ($f_i$):
\begin{eqnarray}
	\sum_i f_i & = & \rho \\
	\sum_i f_i c_i & = & \rho {\bf u}
\end{eqnarray}
Note that the macroscopic drift velocity (${\bf u}$)
is the net result of the particle velocities {$c_i$}
at a particular lattice site.

The post-collision particle distribution function ($f_i^{'}$) is
given by:
\begin{equation}
	f_i^{'}({\bf x},t) = f_i({\bf x},t) + \Omega_i(f_i,..,f_N)
\end{equation}
The function $\Omega_i(f_i,..,f_N)$ describes the collision proces.
The collision function conserves mass and momentum.
This means the following restrictions apply to the collision process:
\begin{eqnarray}
	\sum_i f_i^{'}     & = & \sum_i f_i     \\
	\sum_i f_i^{'} c_i & = & \sum_i f_i c_i
\end{eqnarray}

After the collision the lattice gas particles propagate
to a neighbouring site according to their velocity ($c_i$).
The propagation step is given by:
\begin{equation}
	 f_i({\bf x}+{\bf c}_i,t+1) = f_i^{'}({\bf x},t)
\end{equation}
The propagation step changes neither mass nor momentum of
the individual particles, so the LGA scheme conserves mass and momentum.

LGA can be applied in a broader context
than fluid dynamics. It has been proposed as an alternative to models
with partial differential equations (PDE). Other successful applications
of LGA are natural convection \cite{Argyris93}, reaction-diffusion
      \cite{Ponce93},
and convection-diffusion \cite{Flekkoy93}.
Each particular
transport phenomenon can be set up with an appropriate set of collision rules.

Lattice Gas Automata have some favourable properties over
traditional methods for simulations of models based on PDE's.
With LGA it is particularly simple to implement boundary conditions.
Boundary conditions such as fixed walls are implemented as particles
fixed to a lattice site, which will bounce back the head on colliding
lattice gas particles into the opposite direction from which they came.
The bounce back boundary condition is described by the expression:
\begin{eqnarray}
	f_i'({\bf x},t) & = & f_j({\bf x},t)  \\
	     {\bf c}_i  & = & - {\bf c}_j
\end{eqnarray}
This feature of LGA makes it easy to model the geometry of packagings.

The following reasons yield the choice of LGA as simulation methodology
for modelling the transport phenomena
      in systems of packaged agricultural products:
\begin{itemize}
\item   The methodology is based on simple physical principles. Therefore it is
	easily extendible to other physical phenomena.
\item   Complex geometries are easily modelled by digitizing them
	on the lattice.
\item   Because of their discrete nature,
	LGA can be directly programmed on the computer.
	And because of their simplicity they are easy to program.
\end{itemize}
Compared to the traditional methods LGA demands more computer resources, this
is the price for simplicity.

\section{Transport Phenomena in systems of packaged agricultural products}

To make the physics of conditioning packed agricultural
products more clear, it is first described as a PDE model. This description
is based upon the models by \cite{Bakker67,Beukema80}.
These models consider packed agricultural products as a
porous medium, with the product as the fixed phase and the air as
the flowing phase. Both phases, product and air, can contain heat and vapour.
A forced air flow convects the heat and vapour out of the package.
The medium is assumed to have a homogeneous porosity ($\epsilon$).
The velocity field ($v_a$) is uniform and stationairy.

The complete two-phase PDE-model, describing
     the evolution of the vapour concentration in air
($c_a$), the temperature of the product ($T_p$) and
the temperature of air ($T_a$) in space and time,  is given below:
\begin{eqnarray}
	   \frac{\partial{c_a}}{\partial{t}}
	 +  v_a \frac{\partial{c_a}}{\partial{x}} & = &
	    \frac{( c_{a}^{sat} - c_{a} )}{\tau_{a_c}}
	 +  D \frac{\partial^2 c_a}{\partial x^2}
	 \\
	    \frac{\partial{T_a}}{\partial{t}}
	 +  v_a \frac{\partial{T_a}}{\partial{x}} & = &
	    \frac{( T_p - T_a )}{\tau_{a_T}}
	 +  a \frac{\partial^2 T_a}{\partial x^2}
	 \\
	     \frac{\partial{T_p}}{\partial{t}} & = &
	   - \frac{( T_p - T_a )}{\tau_{p_c}}
	   - \frac{( c_{a}^{sat} - c_{a} )}{\tau_{p_T}}
\end{eqnarray}
Vapour transfer is proportional to the vapour deficit between
the saturated boundary layer around the product ($c_a^{sat}$)
and the air ($c_a$). The boundary layer is kept at
saturation by constant supply of vapour from the water in the plant.
So vapour transfer leads to evaporation of water from the product,
for which heat is extracted from the product itself.

The rates of the heat and vapour transfer are determined by the
time constants ($\{\tau_i\}$), which depend on intrinsic parameters of
the packed product. These parameters are the specific heat capacities
($\rho_i c_{p_i}$), specific heat and vapour resistances
($(\alpha a_{spec})^{-1},(\beta a_{spec})^{-1}$,
porosity ($\epsilon$) and the heat of evaporation ($r$).
The expressions for the time constants are:
\begin{eqnarray}
	    \tau_{a_c} & = & \frac{\epsilon}{\beta a_{spec}}       \\
	    \tau_{a_T} & = & \frac{\rho_a c_{p_a} \epsilon}{\alpha a_{spec}} \\
	    \tau_{p_c} & = & \frac{\rho_p c_{p_p} (1-\epsilon)}
				  {\alpha a_{spec}} \\
	    \tau_{p_T} & = & \frac{r \rho_p c_{p_p}
			 (1-\epsilon)}{\beta a_{spec}}
\end{eqnarray}
Altogether the physics of the conditioning of packed agricultural products
is well known and simple in essence. The difficulty with simulations of
these processes in practice is the formulation of the boundary conditions.
This is quite complicated because of the complex
geometry of the product and the packaging.
This difficulty can be overcome by the use of LGA,
of which one of the strengths is the straightforward handling of
boundary conditions.

\section{Lattice Gas Automata model}

The first goal in this research programme is
      to test the applicability of LGA for the problem
of active conditioning of packed agricultural products. For this test
a one-dimensional (1D) LGA-model has been developed. It is derived from a
convection-diffusion LGA-model \cite{Flekkoy93}. Convection-diffusion
plays a central role in the problem of packaged agricultural products,
as described in the previous section. Before presenting the
complete two-phase LGA-model, the 1D convection-diffusion
LGA model is explained in detail.

\subsection{Convection Diffusion LGA}

Convection-diffusion in 1D can be described as:
\begin{equation}
	   \frac{\partial{\rho}}{\partial{t}}
	 +  u \frac{\partial{\rho}}{\partial{x}} =
	    D \frac{\partial^2 \rho}{\partial x^2}
\end{equation}
This PDE gives the evolution of the mass density of a species ($\rho$) in a
flow
with a velocity field ($u$).

The LGA model maps the problem on a lattice, populated with a lattice gas
consisting of a solvent with tagged particles.
These tagged particles represent the species
which experiences the convection-diffusion. The solvent can be considered as
a 'heat bath', generating the velocity field. The tagged particles
exchange momentum with the 'heat bath' when they collide with the solvent.
Only collisions of the tagged particles with the solvent are considered,
because the density of the tagged particles is assumed to be much smaller
than the density of the solvent.

In the 1D case the lattice is a linear array.
The tagged particles can assume only two velocities ($ c_i = \{ -1,+1 \} $),
allowing the particles to move to the lattice site on
the left or on the right.
The used lattice and velocity set is denoted by D1Q2 in the
terminology of \cite{Qian92}.
The state of a lattice site is determined
by the number of particles ($g_i$) with velocity ($c_{i}$).

The collision rules of the convection-diffusion LGA are
derived from
the discretised relaxation time approximation of the Boltzmann equation,
the Lattice Boltzmann equation (LBE).
Traditionally the Boltzmann equation is used to describe
the collision process
between particles on the microscopic level in the field of
solid state physics and kinetic gas theory \cite{BGK}.

The classical Boltzmann equation with a relaxation time
approximation of the collision process is:
\begin{equation}
	   \frac{\partial{g}}{\partial{t}}
	 +  u \frac{\partial{g}}{\partial{x}} =
	    \frac{( g^{eq} - g )}{\tau}
\end{equation}
Here $g$ is the number of particles in a certain state and $g^{eq}$
is the equilibrium distribution. The relaxation parameter
$\tau$ is dependent on the transport coefficients.

The LGA with collision rules based upon the Lattice Boltzmann
equation have been developed very recently \cite{Benzi92,Qian92}.
This class of LGA is much more efficient in simulating transport
phenomena than earlier LGA models \cite{Frisch87}.

Like the Boltzmann equation, the LBE describes the relaxation of the
particle distribution to an equilibrium distribution ($g_i^{eq}$).
The LBE for 1D convection diffusion is given by:
\begin{eqnarray}
	g_i(x+{\bf c}_i,t+1) & = & g_i(x,t) +           \nonumber      \\
			     &   &  \omega {(g_i^{eq}(x,t) - g_i(x,t))} \\
	g_i^{eq}(x,t) & =  & \frac{\rho}{2} (1 + {\bf u} \cdot {\bf c}_i)
\end{eqnarray}
The relaxation parameter $\omega$ is related to the diffusion coefficient :
$ \omega = {( \frac{1}{2} + D )}^{-1} $. \cite{Benzi92}. The equilibrium
distribution follows from the requirements:
\begin{eqnarray}
      \sum_{i} g_i^{eq} & = & \rho \\
      \sum_{i} g_i^{eq} {\bf c}_i & = & \rho {\bf u}
\end{eqnarray}
That this LGA model leads to convection-diffusive behaviour has been
mathematically derived and checked against experiments \cite{Flekkoy93}.

\subsection{Two-phase LGA}

The two-phase PDE-model as described in section 2 is implemented as a
1D LGA scheme. The air phase is modelled as a lattice gas.
The heat and water vapour of the air are modelled
as tagged lattice gas particles being part of a solvent (the air).

The agricultural product is modelled as particles fixed to the lattice sites.
The fixed particles, representing the agricultural product,
can also contain heat and vapour particles.
The density of fixed vapour particles is related to
the saturation vapour concentration of the boundary layer between air and
product.

The convection diffusion LGA scheme,
      as described in the previous section, is applied to
the heat and vapour lattice gas particles. The interaction between
the two phases, the heat and vapour transfer, is modelled as
a scattering process of the lattice gas with the fixed particles.
During the scattering process heat and vapour particles are exchanged
between the two phases. This scattering process is modelled as an extra
term added to the collision step of the convection-diffusion LGA-scheme.
It is modelled analogous to chemical reactions in the
reaction-diffusion LGA-scheme \cite{Ponce93}.

The distribution of heat and vapour particles
of the air phase at a lattice site are denoted by repspectively $h_i$ and
$v_i$. The densities of
      the heat and vapour lattice gas particles of the air phase,
which can be translated into the macroscopic parameters
of air vapour concentration ($c_a$) and air temperature ($T_a$),
are given by:
\begin{eqnarray}
      \sum_{i} h_i  & = & \rho_h  \sim T_a \\
      \sum_{i} v_i  & = & \rho_v  \sim c_a
\end{eqnarray}
The density of fixed heat particles is denoted by $\rho_q$,
which can be related to the product temperatures $T_p$. $\rho_v^{sat}$ is
the density of the fixed vapour particles which is related to
the saturated vapour concentration, $c_a^{sat}$.
$c_a^{sat}$ is temperature dependent: $c_a^{sat} = c_a^{sat}(T_p)$.

The rescaling of the densities of the
      heat and vapour particles to real units is quite arbitrairy.
      When all $h_i$ and $v_i$ are equal to zero
the situation corresponds to air with a vapour concentration $c_a=0$ and
an air temperature $T_p = T_a = T_{ref}$, a certain reference temperature.

The two-phase LGA-model comprises of two steps, the collision step
and the propagation step. The scheme for both steps are discussed
      in detail below.

\paragraph{{\em Collision and Scattering}}
The collision process is modelled by two collision terms:
$\Omega_i$ and $\Phi_i$.
$\Omega_i$ represents the collisions of the tagged particles with the solvent
and $\Phi_i$ represents the scattering of the tagged particles against the
fixed particles of the product phase. The formulae for the collsion step are:
\begin{eqnarray}
	h_i'(x,t) & = & h_i(x,t) + \Omega_i^h + \Phi_i^h \\
	v_i'(x,t) & = & v_i(x,t) + \Omega_i^v + \Phi_i^v   \\
	\Omega_i^h & = & \omega_h {(h_i^{eq}(x,t) - h_i(x,t))} \\
	\Omega_i^v & = & \omega_v {(v_i^{eq}(x,t) - v_i(x,t))} \\
	\Phi_i^h & = & \frac{\phi_h}{2} { ( \rho_q(x,t) - \rho_h(x,t) )} \\
	\Phi_i^v & = & \frac{\phi_v}{2} { ( \rho_v^{sat}(x,t) - \rho_v(x,t) )} \\
	h_i^{eq}(x,t) & =  & \frac{\rho_h}{2} ( 1 + {\bf u} \cdot {\bf c}_i ) \\
	v_i^{eq}(x,t) & =  & \frac{\rho_v}{2} ( 1 + {\bf u} \cdot {\bf c}_i ) \\
	\rho_h^{sat} & = & \rho_h^{sat}(\rho_q) \\
	\rho_q'(x,t) & = & \rho_q(x,t) + \nonumber \\
		     &   & \phi_{qh} { (\rho_h(x,t) - \rho_q(x,t) )} + \nonumber \\
		     &   & \phi_{qv} { (\rho_v(x,t) - \rho_v^{sat}(x,t) )}
\end{eqnarray}

\paragraph{{\em Propagation}}

During the propagation step the tagged lattice gas particles
move to the neighbouring lattice site, according to their velocity $c_i$.
The propagation is described by:
\begin{eqnarray}
	h_i(x+{\bf c}_i,t+1) & = & h_i'(x,t) \\
	v_i(x+{\bf c}_i,t+1) & = & v_i'(x,t) \\
	\rho_q(x,t+1)        & = & \rho_q'(x,t)
\end{eqnarray}
The reciprocal relaxation times $\omega_h$ and $\omega_v$
are determined by the diffusion
coefficients of the heat and vapour particles ($D_h,D_v$). The relation
$\omega_k = {( \frac{1}{2} + D_k )}^{-1}$ still holds. The relaxation
parameters of the scattering process are related to
      the specific heat capacities,
heat and vapour transfer coefficients. They have to be determined by
appropriate rescaling of the real parameter values.
This must be done because LGA calculate everything in dimensionless numbers.

For the rescaling of convection-diffusion
     the so called Peclet number must be invariant.
For diffusion the Peclet number is $Pe={ L v_a }/{D}$,
      where $L$ is the length scale
(per example the length of the packaging box). For the relaxation
parameters of the scattering process the time constants (12)--(15)
are rescaled accoordingly the time rescaling ($S_t$):
\begin{equation}
	S_t = \frac{L}{v_a} \frac{u}{S}
\end{equation}
Here $S$ is the number of sites of the lattice.

\section{Results and Discussion}

The 1D two-phase LGA-model is tested against
cooling down experiments of packed cut flowers.
The experiments are done with irises packed in a commercially used box.
The initially warm flowers are cooled down by forcing cold air
through the packages, containing the flowers.
The packages have holes in the front and back sides allowing
air flow through the box.

The boxes are used in road transport.
The dimensions of a box are 1.20m by 0.45m by 0.30m.
One box normally contains 18 bunches of 50 irises.
The total weight of the flowers is typically 35 kg.
The flower buds lie either at the front or the back side of the box.
The positioning of the flowers in the box is drawn in figure 2.
Each bunch is wrapped in folio, impermeable to air flow and vapour transport.
The air is ventilated along the length direction of
the box and flows through the bunches.

\begin{figure}[hbt]
\epsfxsize = \hsize
\epsfbox[52 529 545 643]{fig2pos.eps}
\caption{Positioning of flowers inside the package}
\end{figure}

In experiments the flower temperature is measured at several points
      in three cross sections
of the box (at the frontside, in the middle and at the backside of the box,
with respect to the direction of the air flow).
      The experimental results in figure 3 are obtained
after averaging over the points in the cross sections.

The two-phase LGA-model is tested against these experimental results.
For most factors determining the time constants (12)--(15)
the values are known from experiments or from literature.
Only the resistances to heat and vapour transfer have to be estimated.
The heat resistance is estimated from earlier experiments.
For the most part the resistance of vapour transfer is related to the heat
resistance,
this is the Lewis relation \cite{Ashrae69}. Added to this vapour resistance
is a small contribution of resistance of the plant tissue. The initial
temperature of the air and the product ($18 ^oC$)  and the relative humidity
of the air ($R.H. = 90\%$) are taken as initial conditions for the LGA scheme.
Boundary conditions have to be maintained at the inlet and outlet of the box.
There is an inflow of air of 0.1 $m/s$, $3^oC$ and $80\% R.H.$
There is an outflow of air of 0.1 $m/s$
with temperature and vapour content of the air in the last cross section.

\begin{figure}[hbt]
\epsfxsize = \hsize
\epsfbox[24 369 588 768]{fig3exp.eps}
\caption{Experimental results versus simulation results. Experiments are
	 done during cooling down of packaged cut flowers}
\end{figure}

The LGA model has still two free parameters: the lattice size ($S$) and
the drift velocity ($u$). With these parameters one can set the preferred
resolution in time and space. The limits for these parameters are set by
the requirements that the time step ($dt$) must be smaller than the retention
time of the air flow at one lattice site, which means that the drift velocity
($u$) must be smaller than the propagation velocity ($c=1$)
of the lattice gas particles: $u \ll 1$.

The simulation results are fitted to the experimental
results using fine tuning of the heat resistance and the
vapour resistance of the plant tissue. The simulation results are
also shown in figure 3 as solid curves.
Comparision with the experimental results indicate that
      the 1D two-phase LGA model, is able to simulate
the average behaviour of packed agricultural products.

Future development of the two phase LGA
      model will be the extension to higher dimensions
(at least 2D). This extension is needed in order to
      simulate the spatial variation of
temperature orthogonally to the air flow direction, for which significantly
differences have been found in the experiments on packed irises. These
variations in temperature are likely
      a consequence of the location of the holes in the package
and the way of stacking the bunches of flowers inside the box. So in order to
meet the obejctive of this research programme,
the optimization of the package design for the achievement of
improved product quality, a more detailed model is needed. This model
must explain the differences in product behaviour induced by the packaging
design.

\bibliographystyle{named}
\bibliography{d:esm}

\end{document}